\documentclass[twocolumn]{openjournal}

\usepackage{xcolor}
\usepackage{textgreek}
\usepackage[utf8]{inputenc}
\usepackage[english]{babel}
\usepackage{amsmath}
\usepackage{graphicx}
\usepackage{floatrow}

\usepackage{hyperref}
\hypersetup{
    unicode, 
    colorlinks=true,
    linkcolor=linkcolor,
    citecolor=linkcolor,
    filecolor=linkcolor,
    urlcolor=linkcolor,
}
\usepackage{color,colortbl}
\definecolor{linkcolor}{rgb}{0.0,0.3,0.5}
\DeclareGraphicsExtensions{.bmp,.png,.jpg,.pdf}
\usepackage{verbatim}
\usepackage[normalem]{ulem}
\usepackage{orcidlink}
\usepackage{soul}

\graphicspath{ {./figs/} }

\begin{document}
\title{Maximizing Ariel's Survey Leverage for Population-Level Studies of Exoplanets}

\author{Nicolas B.\ Cowan\orcidlink{0000-0001-6129-5699}}
\email{nicolas.cowan@mcgill.ca}
\affiliation{Department of Physics, McGill University, 3600 rue University, Montréal QC H3A 2T8, Canada}
\affiliation{Department of Earth \& Planetary Sciences, McGill University, 3450 rue University,
Montréal, QC H3A 0E8, Canada}

\author{Ben Coull-Neveu\orcidlink{0009-0000-2192-6003}}
\email{benjamin.coull-neveu@mail.mcgill.ca}
\affiliation{Department of Physics, McGill University, 3600 rue University, Montréal QC H3A 2T8, Canada}

\begin{abstract}
    ESA's Ariel mission will be uniquely suited to performing population-level studies of exoplanets. Most of these studies consist of quantifying trends between an Ariel-measured \emph{a posteriori} quantity, $y$, and an \emph{a priori} planetary property, $x$; for example, atmospheric metallicity as inferred from Ariel transit spectroscopy vs.\ planetary mass. 
    The precision with which we can quantify such trends depends on the number of targets in the survey and their variance in the \emph{a priori} parameter. We define the leverage of a survey with $N$ targets as $L \equiv \sqrt{N} \, \text{stdev}(x)$ and show that it quantitatively predicts the precision of population-level trends. The target selection challenge of Ariel can therefore be summarized as maximizing $L$ along some axes of diversity for a given cumulative observing time. To this end, we consider different schemes to select the mission reference sample for a notional three year transit spectroscopy survey with Ariel. We divide the exoplanets in the mission candidate sample into logarithmic classes based on radius, equilibrium temperature, and host star temperature. We then construct a target list by cyclically choosing the easiest remaining target in each class.   
    We find that in many cases the leverage is greatest for a single class: dividing planets into multiple classes increases the diversity of targets, but reduces their numbers. The leverage on a single axis of diversity can be increased by dividing that axis into many planet classes, but this sacrifices leverage along other axes of diversity. 
    We conclude that a modest number of classes ---possibly only one--- should be defined when selecting Ariel targets. Lastly, we note that the statistical leverage of the Ariel transit survey would be significantly increased if current candidate planets were confirmed. This highlights the urgency of vetting and confirming the easiest transmission and emission spectroscopy targets in the Ariel mission candidate sample.
\end{abstract}

\begin{keywords}
    {Surveys, Planets and satellites: atmospheres, Methods: statistical}
\end{keywords}

\maketitle

\section{Introduction}
\label{sec:intro}
Exoplanets are harder to study than solar system worlds, but they are more numerous. Exoplanets present a unique opportunity to extend comparative planetology to populations of dozens or hundreds of planets \citep[e.g.,][]{2011ApJ...729...54C,2016Natur.529...59S,2019ApJ...887L..20W,2021NatAs...5.1224M, 2021MNRAS.504.3316B, 2021AJ....162...36W,2022ApJS..260....3C, 2023ApJS..269...31E,2025AJ....169...32D,2024arXiv241206573P}. The European Space Agency's Ariel mission will perform an unprecedented uniform survey of exoplanet atmospheres and hence quantify population-level trends \citep{2018ExA....46..135T,2019PASP..131i4401Z}. 
 
Which exoplanets should Ariel observe during its four year primary mission? This is an urgent question because the Ariel target list must be finalized one year prior to launch, scheduled for late 2029 \citep{tinetti2021arielenablingplanetaryscience}.  Many of the most promising potential Ariel targets have yet to be confirmed, and doing so can take years. Astronomers therefore need to know which exoplanet candidates and host stars should be prioritized for stellar characterization \citep{2022ExA....53..473D}, vetting, radial velocity, and ephemeris maintenance \citep{2020PASP..132e4401Z,2022ExA....53..547K,2025arXiv250803801B}.

\cite{2022AJ....164...15E} defined a mission candidate sample (MCS) for the Ariel mission based on NASA's Exoplanet Archive \citep{2025arXiv250603299C} and ArielRad \citep{2020ExA....50..303M,2025ExA....59....9M}, building on the work of \cite{2018ExA....46...67Z}. The MCS includes both confirmed and candidate planets, is regularly updated and is available on GitHub.\footnote{\href{https://github.com/arielmission-space/Mission_Candidate_Sample}{Ariel MCS GitHub}} We adopt the MCS from 9 July 2024 and consider how different selection criteria impact the quality of Ariel science. Since the focus of the mission will be population-level trends in exoplanet atmospheres, Ariel should obtain high signal-to-noise ratio (SNR) observations of a large number of diverse targets. Target selection is non-trivial, however, because there are trade-offs between SNR, number, and diversity of targets. 

The axes of diversity needed for Ariel science depend on the specific science questions one is trying to answer and the kind of observations considered, e.g., transmission vs.\ emission spectroscopy. 
Previous authors have considered diversity in planetary radius $R_{\rm p}$, equilibrium temperature $T_{\rm eq}$, and sometimes stellar effective temperature $T_{\rm eff}$ \citep{2022AJ....164...15E,2023AJ....165...14B,2024AJ....167..233H}. We adopt these three axes of diversity in the current paper, with the understanding that our approach could be adapted to other axes of diversity.

We consider a notional Ariel transit spectroscopy survey spanning three years of cumulative observing time.  Moreover, we adopt a uniform Tier 2 SNR for all targets. Tier 2 observations detect the atmospheric scale height of the planet's atmosphere in transit at $\geq 7\sigma$ at spectral resolution of $R\approx10,50,10$ in Ariel's NIRSpec, AIRS channels 0 and 1 \citep{2019AJ....157..242E}. This data quality should be sufficient for spectral retrieval \citep{2022ExA....53..447B}. 
Moreover, the Tier 2 transit survey is expected to make up the bulk of the Ariel mission, and the assumption of uniform SNR on transit spectral features simplifies the target selection problem.
  
Our challenge is to select the subset of the MCS that should be observed in a notional three year Ariel Tier-2 transit survey to maximize the science output. The rest of this paper is organized as follows: in \S2 we define the figure of merit to be optimized along each axis of diversity. In \S3 we describe the class structures and selection schemes we test to enforce target diversity. We present our results in \S4 and discuss their implications in \S5.   

\section{Diversity and Leverage}
\label{sec:leverage}
\begin{figure}[ht]
    \centering
    \includegraphics[width=1.0\linewidth]{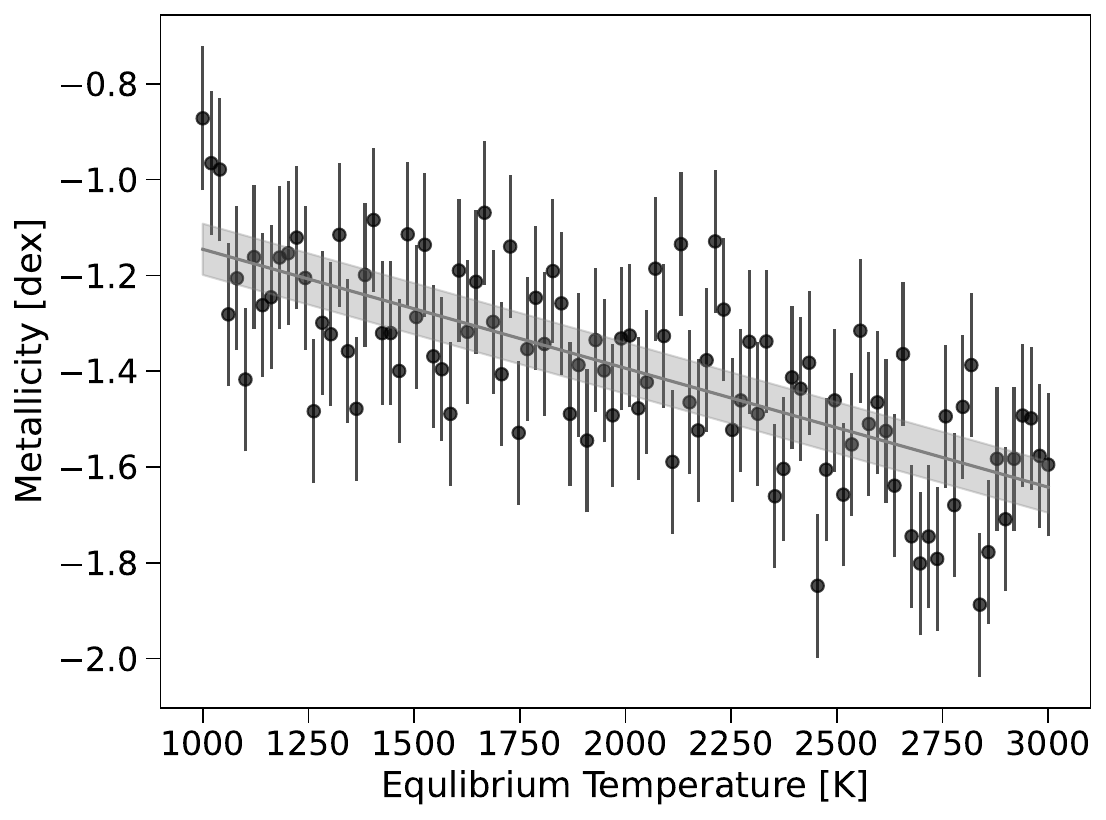}
    \caption{Example science case for the Ariel transit spectroscopy survey. Each black dot represents a hypothetical planet with a precisely measured equilibrium temperature and uniform uncertainties on the retrieved atmospheric metallicity (motivated by the uniform signal-to-noise of Ariel Tier 2 transit spectra). The gray line and band shows the best-fit linear trend and its $1\sigma$ confidence interval. The Ariel target-selection problem for this science case amounts to choosing the planets that minimize the uncertainty on the slope of the trend line, for a given amount of cumulative observing time.}
    \label{fig: metallicity vs Teq}
\end{figure}

Most Ariel science can be summarized as quantifying population-level trends in atmospheric properties. The simplest trends are linear: $y = mx+b$, where $y$ represents an \emph{a posteriori} planetary property measured with Ariel and $x$ is a planetary parameter known \emph{a priori} \citep[e.g.,][]{2016ApJ...831...64T,2019ApJ...887L..20W,2024SSRv..220...61S}. 
Figure \ref{fig: metallicity vs Teq} shows a notional (log-)linear trend between atmospheric metallicity and planetary equilibrium temperature; in \S\ref{sec:discussion} we discuss non-linear trends. 

Targets should therefore be selected to maximize the precision with which this trend can be estimated. 
To simplify the problem, we neglect the uncertainty on the \emph{a priori} parameter, $\sigma_x = 0$, since in practice these are usually negligible. Moreover, we assume homogeneous uncertainties in the Ariel-measured quantities, $\sigma_y$, which is motivated by the uniform SNR of the Ariel Tier 2 survey \citep[e.g., the expected uncertainty on H$_2$O abundance from retrievals of Tier 2 Ariel transmission spectra of gas giants is $0.10\pm0.07$ dex:][]{2023RASTI...2...45C,2025arXiv250713446D}. The uncertainty on the intercept is essentially the error on the mean: $\sigma_b=\sigma_y/\sqrt{N}$. Minimizing $\sigma_b$ would simply mean observing as many targets as possible, regardless of their diversity. The slope of the trend is more interesting, so the Ariel target selection problem boils down to choosing the set $\{x_i\}$ that minimizes $\sigma_m$.   

We expect the precision of a linear trend to improve as the square root of the number of planets, $\sqrt{N}$. The diversity of the sample in $x$ can be  quantified via the standard deviation, stdev($x$). We therefore hypothesize that the precision with which one can constrain a population-level trend depends on the \emph{survey leverage} in $x$:
\begin{equation}
    L \equiv \sqrt{N}\, \text{stdev}(x) = \sqrt{\sum_{i=1}^N(x_i-\bar x)^2},
\end{equation} 
where $\bar x$ is the mean of $\{x_i\}$. In other words, a population's leverage is the quadrature sum of deviations from its mean. Statisticians fret about a few data with too much leverage throwing off fits \citep{10.1214/aos/1176342811,10.1214/ss/1177013622}, and this could indeed be a problem for comparative planetology involving only a few targets. For a survey of hundreds of planets, however, aggregate leverage is desirable.

In fact, for least squares fitting, the uncertainty on the slope of a linear function is inversely proportional to the survey leverage (details in Appendix):
\begin{equation}
    \sigma_m = \frac{\sigma_y}{\sqrt{\sum(x_i-\bar x)^2}} = \frac{\sigma_y}{L}.
\end{equation}  
We confirmed this analytic relation via numerical tests. 
In addition to being intuitive, the survey leverage quantitatively predicts the precision of a linear trend. 

The Ariel target selection problem can therefore be recast as: from the $N_{\rm MCS}$ potential targets in the mission candidate sample, choose the $N$ planets to be observed such as to maximize the leverage, $L$, in a given cumulative observing time. \cite{2023AJ....165...14B} developed a figure-of-merit to choose individual targets to tackle a focused science question, but such a strategy does not work for maximizing the survey leverage, which is a combinatorial problem. Given the number of planets in the MCS, $N_{\rm MCS}\approx 3000$, and the number of planets Ariel is expected to observe in Tier 2, $N\approx 600$, the number of permutations given by the binomial coefficient is astonishing: $N_{\rm MCS}!/(N!\,(N_{\rm MCS}-N)!) \approx 10^{650}$.  It is impossible to exhaustively consider each and every combination of targets, so we proceed with simple heuristics.   

\section{Planetary Classes}
The standard way to increase the diversity of a sample of exoplanets is to define different planetary classes, e.g., hot Jupiters, warm sub-Neptunes, lava planets. When selecting targets, one chooses the easiest targets of each class. This is the “best-in-class” approach advocated by \cite{2022AJ....164...15E} and \cite{2024AJ....167..233H}, although “easiest-in-class” would be more accurate. There are two subtleties to this scheme: exactly how the planetary classes are defined, and how the targets are selected.  

\subsection{Class Structures}
The most popular approach to date has been to define classes by eye, sometimes based on an understanding of exoplanetary atmospheres, a proclivity for round numbers, or the historical definitions of stellar spectral types. \cite{2024AJ....167..233H} defined such classes for the axes of planetary radius and equilibrium temperature, while \cite{2022AJ....164...15E} and \cite{2023AJ....165...14B} also included the spectral type of the host star as a third axis of diversity. Such classes are not equally sized either linearly or logarithmically, so it is difficult to test the impact of increasing the number of classes. Since we specifically want to test how the number of classes impacts survey leverage, we do not consider ad hoc classes.  

\begin{figure}[htb]
    \centering
    \includegraphics[width=\linewidth]{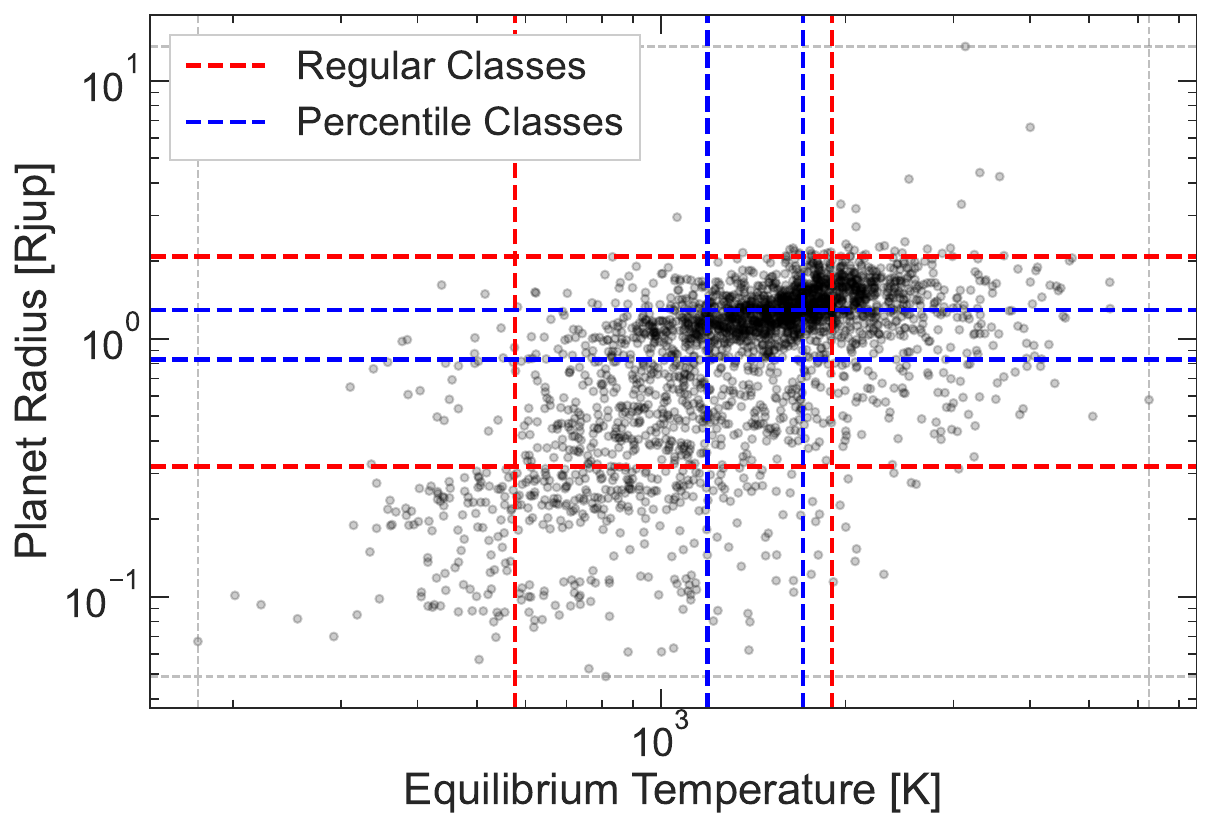}
    \caption{An example $3\times3$ logarithmic class structure for two axes of diversity: planetary radius, $R_{\rm p}$, and equilibrium temperature, $T_{\rm eq}$. Gray dots show the Ariel mission candidate sample from \cite{2022AJ....164...15E}, while gray dotted lines denote the extrema of the sample. The dashed red lines show the class boundaries for the regular classes (equally dividing the parameter space bounded by the extrema), while the blue dashed lines show the boundaries between percentile classes (equally dividing the planets based on a ranked list).}
    \label{fig: classes}
\end{figure}

Alternatively, we can subdivide each axis of diversity into equal-size classes (red lines in Figure \ref{fig: classes}). The boundaries between these \emph{regular classes} are given by:
\begin{equation}
    x_j = x_{\rm min} + j(x_{\rm max}-x_{\rm min})/N_{\rm class},
\end{equation}
where $x_{\rm min}$ and $x_{\rm max}$ are the minimum and maximum value of a given \emph{a priori} parameter in the overall sample, classes are indexed with $j \in \{1,2,\ldots N_{\rm class}\}$, and $N_{\rm class}$ is the number of classes along the $x$-axis. The classes tend to make more sense in logarithmic space, e.g., $x=\log T_{\rm eq}$, as shown in Fig \ref{fig: classes}. That said, our results are essentially the same when we work in linear space.

Regular classes suffer the problem that certain classes are sparsely populated.  In fact, if one is splitting the planets into classes in multiple dimensions of diversity, it is likely that many of the classes will be empty. Thinly-populated classes present challenges for the selection process described in the next section. One can instead define \emph{percentile classes} (blue lines in Figure \ref{fig: classes}).  When only considering a single axis of diversity, this class structure ensures equal representation in each class: by definition there are as many planets between the 25$^{\rm th}$ and 50$^{\rm th}$ percentile as between the 50$^{\rm th}$ and 75$^{\rm th}$, regardless of where those boundaries lie in physical units. Unlike regular classes, percentile classes are the same for linear or logarithmic axes of diversity. When considering multiple axes of diversity, the sparsely-populated classes are fewer than with regular classes.  

Regardless of how the classes are defined, the limiting case is a single all-encompassing class, $N_{\rm class}=1$, which we refer to as the \emph{one-room schoolhouse}. This limit is of interest because it necessarily leads to the most planets being observed in a given amount of time. Since the number of targets contributes to the leverage of a sample, the one-room schoolhouse is often near-optimal, as we will see below.  

\subsection{Selecting Targets}
Dividing planets into classes only increases the diversity of the sample if targets are chosen from a variety of classes. The simplest scheme is to sequentially cycle through each class, always selecting the easiest remaining target in that class. In practice, one may run out of planets in a given class before the end of the survey, or a class might have been empty to begin with. In such a case, we simply skip that class in subsequent cycles. 

Alternatively, one could run the easiest-in-class selection for a set number of cycles before switching to a one-room schoolhouse for the remainder of targets. This is essentially what \cite{2022AJ....164...15E} adopted, with two cycles. Their resulting mission reference sample is therefore the union of two distinct target lists: an easiest-in-class scheme with ad hoc classes and a one-room schoolhouse. The motivation for their strategy was that some classes had few planets that are easy to observe; if the cyclical selection were to continue indefinitely, then very challenging targets would have to be observed, which would use a prohibitively large fraction of the remaining mission time. Another way of thinking about this problem is that if you quickly run out of suitable targets in a class, then you probably defined the class too narrowly given the potential targets for the survey in question. We are interested in the simplest selection schemes, so we do not limit the number of cyclical selection rounds. In other words, we only consider monolithic target lists selected via a single scheme.

The targets within each class are ranked in order of their required observation time to achieve Tier 2 quality. This maximizes the number of targets that can be observed within a given amount of time. The time allocated to each observation is $3\times$ that planet's transit duration, providing a generous baseline.\footnote{The Ariel Consortium usually adopts $2.5\times$ the transit duration \citep[][]{2022ExA....53..807M}, but we neglect slew time and other scheduling realities, so we conservatively adopt $3\times$ the transit duration.} Each planet is re-observed until it satisfies the Tier 2 criteria. All the relevant numbers are conveniently listed in the \cite{2022AJ....164...15E} MCS on GitHub.

\section{Results}

\begin{figure*}[htb]
    \centering
    \includegraphics[width=\textwidth]{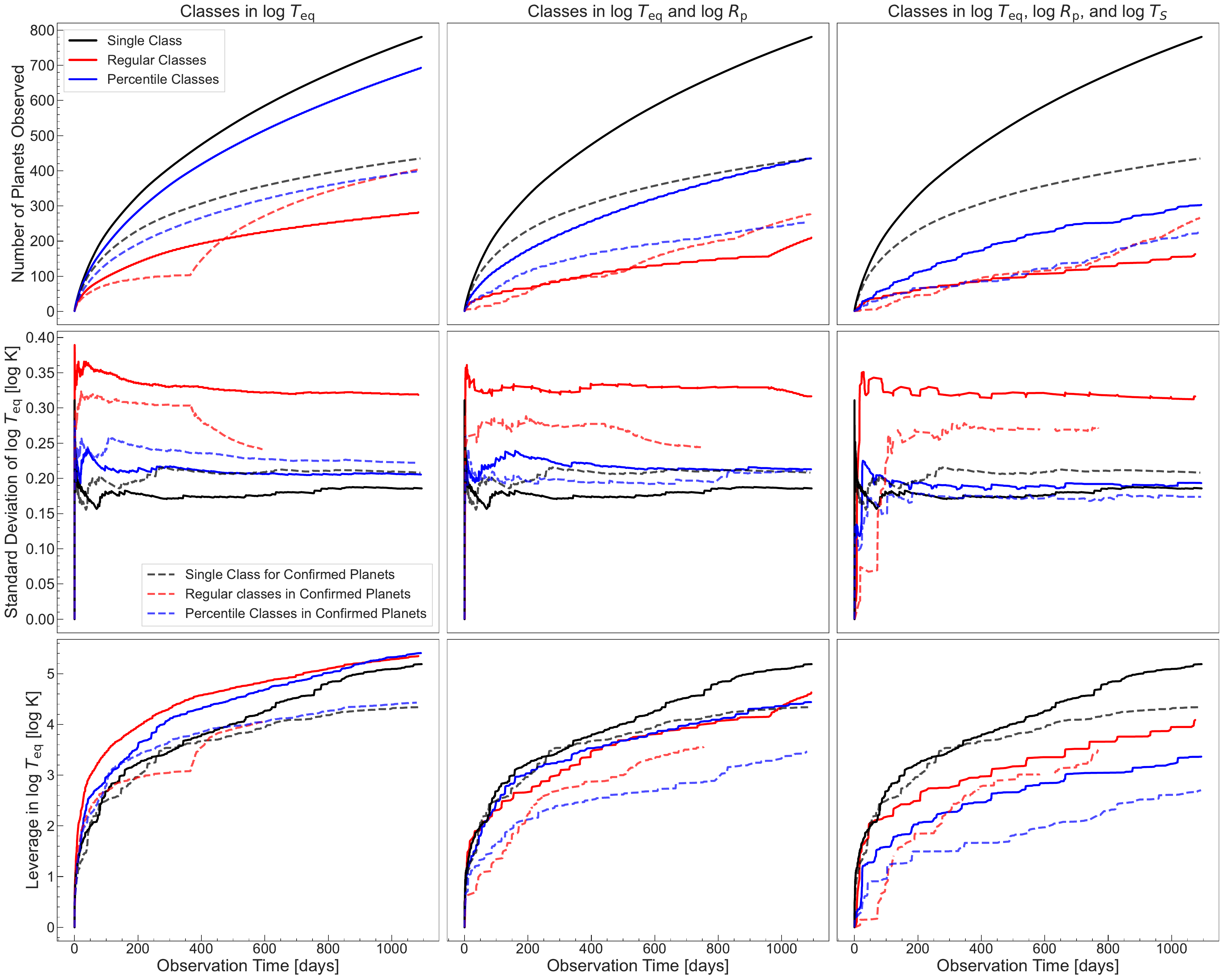}
    \caption{Number of observed planets (\emph{top}), standard deviation of the population (\emph{middle}), and leverage on equilibrium temperature (\emph{bottom}) as a function of total observing time. From left to right we consider an increasing number of axes of diversity: only equilibrium temperature (\emph{left}), also planet radius (\emph{middle}), and also stellar effective temperature (\emph{right}), always with three classes per axis. The central column corresponds to the $3\times 3$ logarithmic class system shown in Figure \ref{fig: classes}. Surveys using only confirmed planets (dashed lines) observe 44\% to 71\% fewer targets and provide 15\% to 27\% less leverage. Dividing planets into regular classes improves leverage when only considering one axis of diversity (left column), but this advantage disappears ---and becomes a disadvantage--- when considering more axes.
    }
    \label{fig:leverage_with_axes}
\end{figure*}

We now present the survey leverage for many hypothetical three year Ariel transit surveys, in which we vary the number of axes of diversity and the number of classes for each axis. Figure \ref{fig:leverage_with_axes} shows the impact of increasing the number of axes of diversity while applying 3 classes to each dimension. 
We divide the planets into logarithmic classes in equilibrium temperature, planetary radius, and stellar temperature. 
The middle column corresponds to the $3\times 3$ class system shown in Figure \ref{fig: classes}. 
Compared to the one-room schoolhouse, the number of targets observed after three years is 64\% lower for 3 regular classes and 11\% lower for 3 percentile classes, 73\% lower for $3\times3$ regular classes and 44\% lower for $3\times 3$ percentile classes, and 79\% lower for $3\times3\times3$ regular classes and 61\% lower for $3\times 3 \times 3$ percentile classes. Dividing planets into classes necessarily reduces the number of targets that can be observed in a given time.  Percentile classes have less of an impact than regular classes, both on the number of targets observed, and on their diversity. Planet classes, especially regular classes, increase sample diversity, but the one-room schoolhouse offers the most survey leverage when considering multiple axes of diversity. We obtained similar results for leverage on $R_{\rm p}$, and for both linear and logarithmic axes: dividing planets into classes increases sample diversity, but the fewer planets results in less leverage than the one-room schoolhouse. 

\begin{figure*}[htb]
    \centering
    \includegraphics[width=\linewidth]{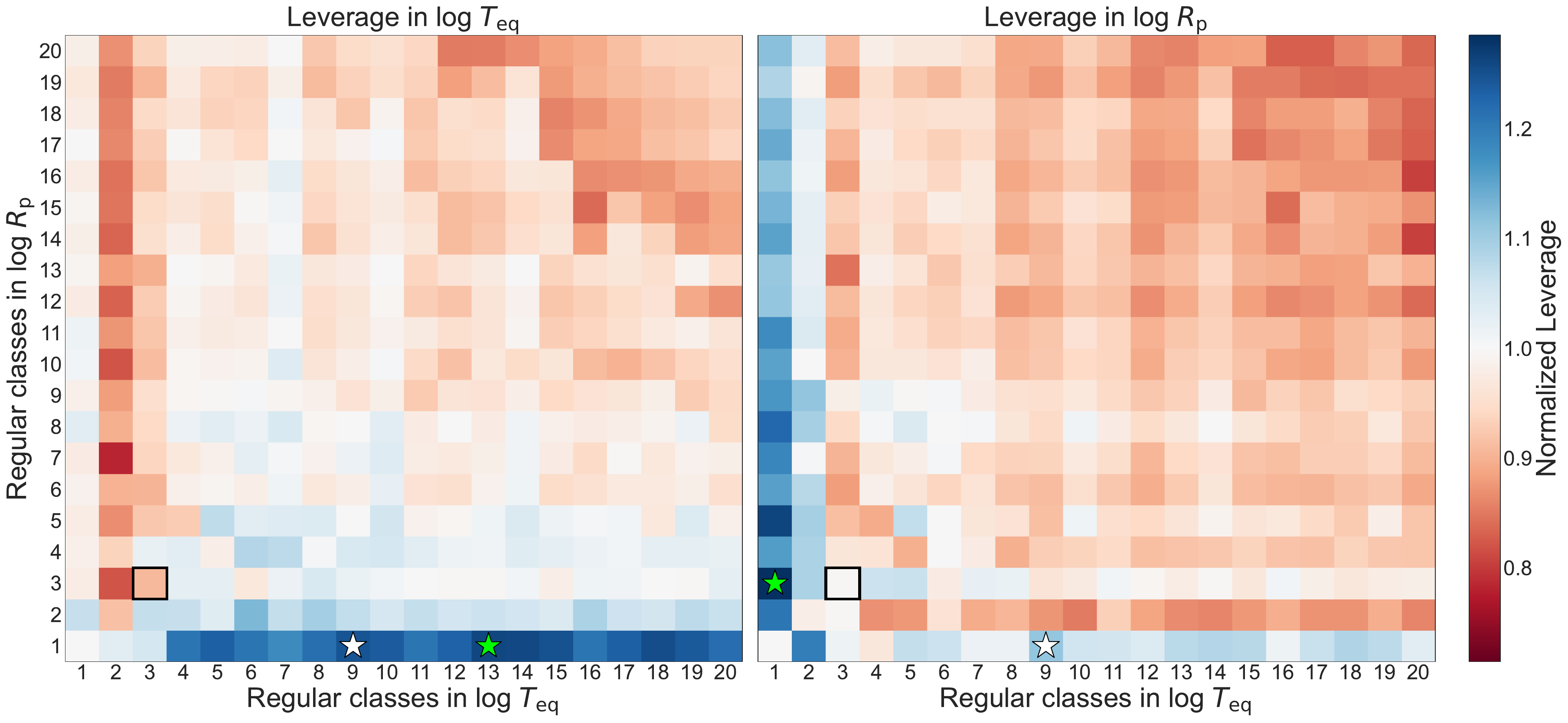}
    \caption{Leverage in equilibrium temperature (\emph{left}) and planetary radius (\emph{right}) for 2D regular classes in log space. For each panel the leverage is normalized to that of the one-room schoolhouse (bottom left corner of each panel). The green star in each panel highlights the class system yielding greatest leverage in that axis. The sum of the normalized leverages in \emph{both} $T_{\rm eq}$ and $R_{\rm p}$ is shown by a white star. The $3\times 3$ class structure used in previous figures is highlighted by a black box. The best leverages achievable are $\sim$25\% better than that of the one-room schoolhouse (light blue), but most class structures lead to somewhat worse leverage (shades of red).}
    \label{fig: normalize leverage heatmaps}
\end{figure*}

Figure \ref{fig: normalize leverage heatmaps} shows how the final leverage after 3 years of cumulative observations depends on the number of regular classes used in a two-dimensional logarithmic class system for equilibrium temperature and planetary radius. The leverages were normalized to the value achieved by the one-room schoolhouse, i.e., the $1\times1$ class system at the bottom left of each panel.  
Leverage on $T_{\rm eq}$ is maximized for 13 classes in equilibrium temperature and a single class in planetary radius ($13\times1$).  Conversely, leverage on $R_{\rm p}$ is maximized for a single class in equilibrium temperature and 3 classes in radius ($1\times3$). Compromise is needed to obtain high leverage in both $T_{\rm eq}$ and $R_{\rm p}$. The sum of leverages in the two dimensions of diversity is maximized for $9\times1$ classes. When we calculated the leverage in linear space or with a slightly different signal-to-noise convention, our results were qualitatively the same, but the optimal class structures shifted a bit. 

We also simulated target selection for the three axes of diversity shown in the right column of Figure \ref{fig:leverage_with_axes}, but varying the number of classes along each axis of diversity.  The axes are ordered as $T_{\rm eq} \times R_{\rm p} \times T_{\rm s}$. This is harder to plot, but can be summarized as follows: the best leverage for equilibrium temperature occurs in the $13\times1\times1$ class system. For planetary radius, it occurs in the $1\times3\times2$ class system. For stellar temperature, it occurs in the $1\times1\times8$. If the relative leverages for each dimension are summed, then the best overall leverage is obtained with the $1\times1\times5$ class system, i.e., a single class in $T_{\rm eq}$ and $R_{\rm p}$, with five classes in $T_{\rm s}$. In other words: dividing planets into classes along some axis of diversity can improve the leverage for that axis, but usually at the cost of reducing the leverage along other axes.

\section{Discussion}
\label{sec:discussion}
\subsection{Leverage not tokenism}
We have shown that increasing the diversity of Ariel targets by defining classes of planets does not necessarily maximize survey leverage. Observing the easiest targets in a one-room schoolhouse provides good leverage and may be close to the optimal solution when considering multiple axes of diversity.  Hard targets in corners of parameter space should not be included in the target list unless they provide more leverage for survey science than a larger number of easier targets. If a hard-to-observe target is interesting on its own, then it should be observed with the James Webb Space Telescope, which has greater photon collecting area than Ariel.

Dividing potential targets into classes along some axis generally increases the diversity of the resulting sample along that axis, but necessarily reduces the number of targets that can be observed in a fixed amount of time: the additional constraints lead to selection of harder targets and hence more observing time per target, on average.  The severity of this time penalty is greater if there are more poorly-populated classes of planets because there is more risk of being forced to choose very time-consuming targets. On the other hand, evenly-populated classes based on percentiles tend to have less impact on the diversity of the sample, but also less impact on the number of targets observed. When many axes of diversity are considered, regular classes seem to provide better leverage than percentile classes.

\subsection{Survey size matters}
The leverage of a target list predicts the precision achievable for linear trends. Some trends will instead exhibit a knee, step function, or non-monotonic behaviour. Regardless of the functional forms explored, such population-level trends would be best quantified with hierarchical modelling, which itself benefits from more planets \citep{2022MNRAS.509..289K,2022AJ....163..140L}. We leave a quantitative study of these behaviours for future research, but we speculate that survey size and leverage will remain useful figures of merit in general.

\begin{figure*}[htb]
    \centering
    \includegraphics[width=0.49\linewidth]{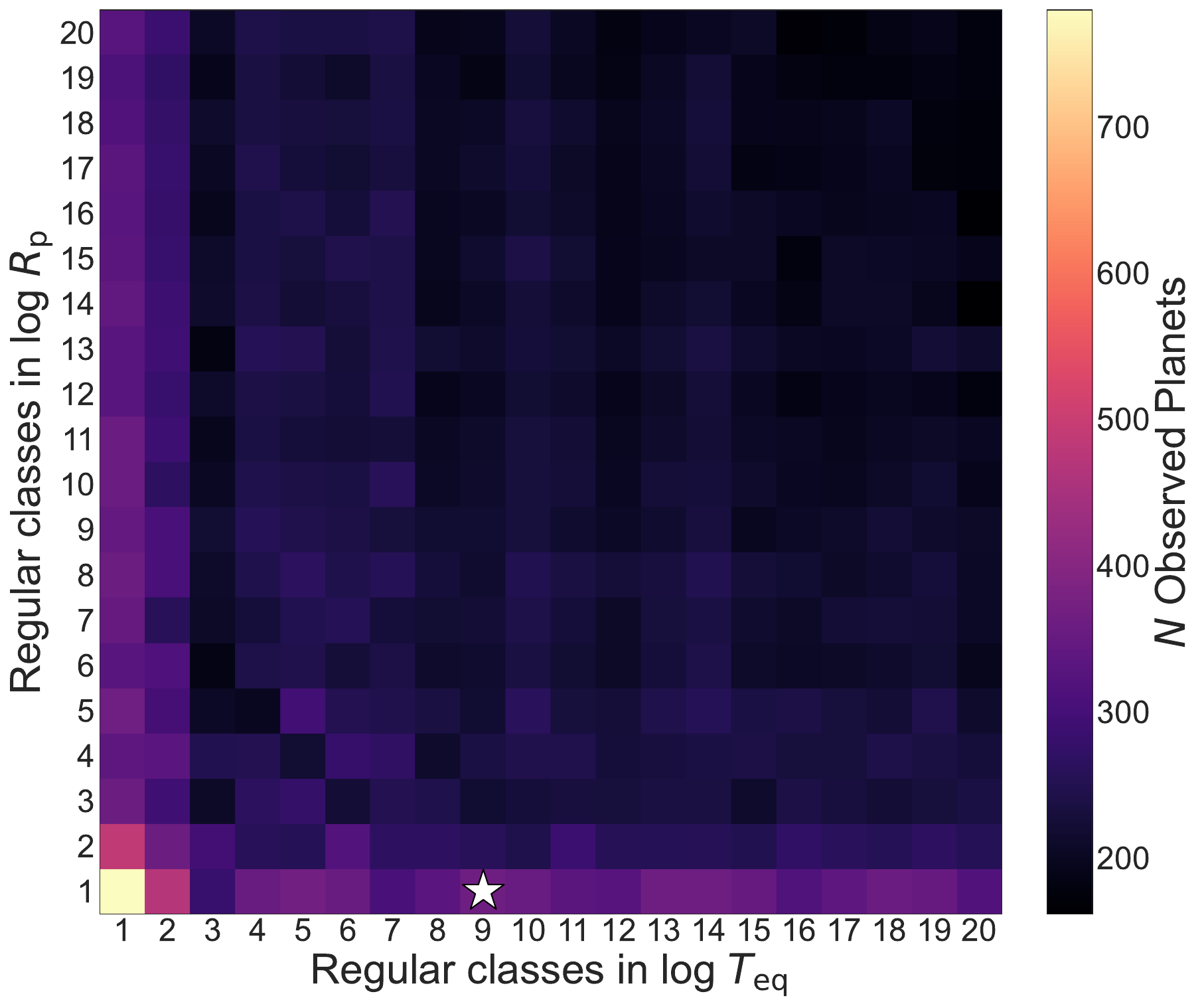} \includegraphics[width=0.49\linewidth]{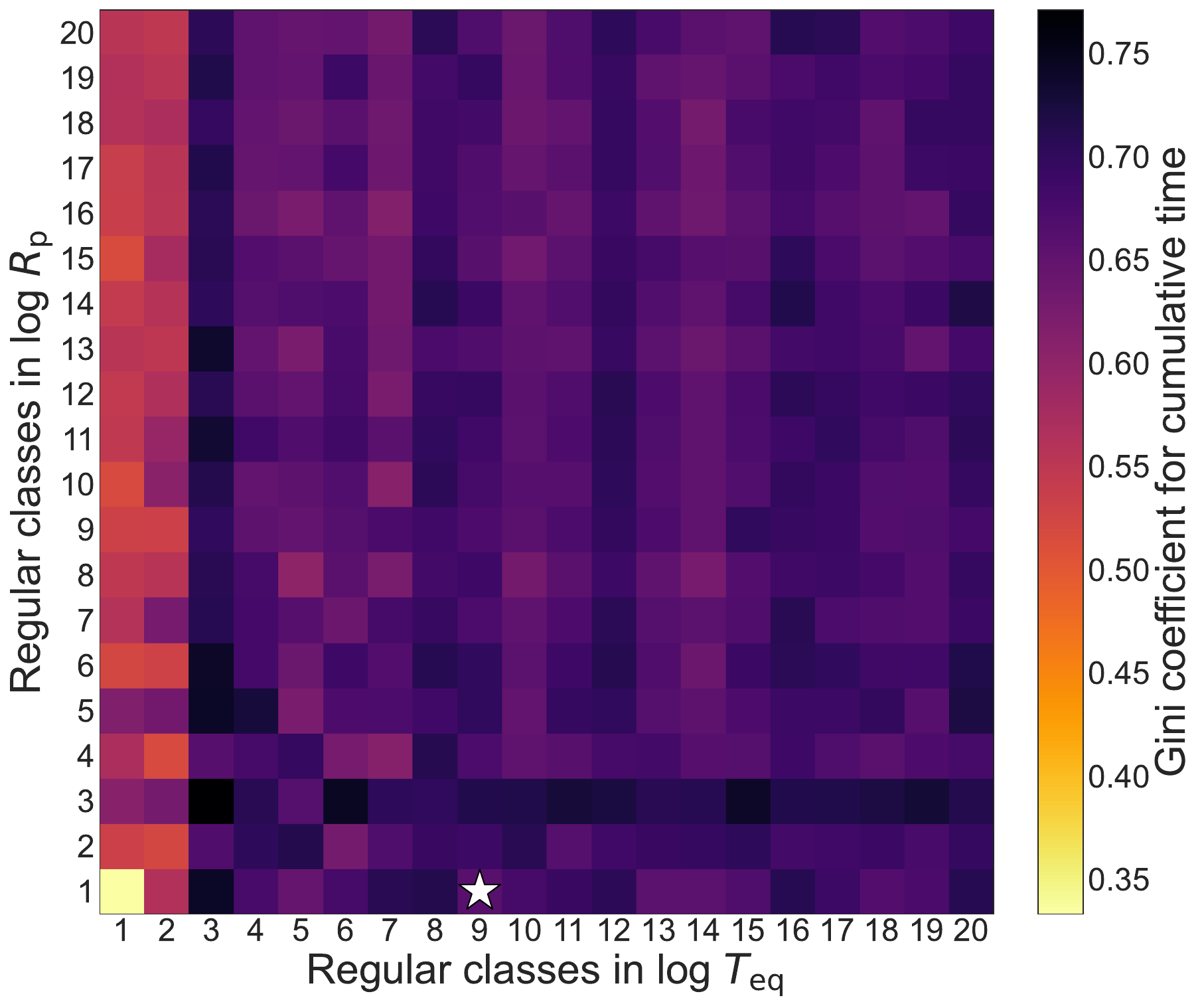}
    \caption{Number of planets observed (\emph{left}) and resulting Gini coefficient (\emph{right}) for a notional 3 year Ariel transit survey. The white star highlights the best overall leverage. The one-room schoolhouse yields the most targets, with other scenarios resulting in 30--60\% fewer planets and Gini coefficients 50--100\% worse. The one-room schoolhouse is the most equitable option and in general the Gini coefficient is lower for class structures that lead to more planets being observed.}
    \label{fig: observed planets heatmap}
\end{figure*}

More broadly, not all Ariel science cases will consist of measuring linear trends in a single \emph{a posteriori} atmospheric property as a function of a single \emph{a priori} parameter. Some population-level trends could be sharp and hence benefit from a concentration of targets in some region of parameter space, e.g., the super-Earth to mini-Neptune transition, or the Cosmic Shoreline \citep[][]{2025arXiv250702136B,2025arXiv250808253I}. Indeed, studying many planets with indistinguishable  \emph{a priori} parameters would be an excellent way to assess the importance of stocasticity in planetary evolution  (D.\ Turrini, priv.\ comm.). Alternatively, we may look for trends between two \emph{a posteriori} parameters (L.\ Mugnai, priv.\ comm.). None of these scenarios benefit from sample leverage as we have defined it, but they all benefit from having more targets. Given a choice between different target lists with comparable leverage, we should therefore favour those with the most targets. In Figure \ref{fig: observed planets heatmap} we show the number of targets for the same 2D classes shown in Figure \ref{fig: normalize leverage heatmaps}. Most class structures reduce the planets observed by more than 70\% compared to the one-room schoolhouse.  The 2D compromise of 9 classes in equilibrium temperature and a single class in planetary radius has 53\% fewer planets. 

\subsection{Scheduling matters}
We have neglected the complication of ephemerides and Ariel's limited field of regard. The curves shown in Figure \ref{fig:leverage_with_axes} presume that Ariel observes the easiest targets first, but in practice scheduling will have to account for which planets are visible and about to transit at each point in time. These complications make the scheduling of Ariel observations non-trivial \citep{2022ExA....53..807M}, so fewer planets can be observed in a given cumulative observing time than the hypothetical numbers presented in this study.

The difficulty of scheduling a target scales as the ratio of transits needed to transits available \citep{2022ExA....53..807M}.  The available transits of a given target depend on its orbital period and location in the sky, which we do not attempt to model in this study. We hypothesize, however, that it will be more challenging to schedule the Ariel survey if targets are divided into classes of orbital period or related quantities (semi-major axis or equilibrium temperature), since longer-period planets will necessarily have fewer available transits.

When designing a large survey, one would like the time spent on each target to be comparable. We quantify the disparities in time-on-target via the Gini coefficient \citep{gini1912variabilita}. In the right panel of Figure \ref{fig: observed planets heatmap} we show the Gini coefficients for the same 2D class structures shown in the previous figures. The observing time disparity is minimized in the one-room schoolhouse: observing the easiest targets results in the most equitable allocation of telescope time.  The scenarios with the highest Gini coefficients also have the fewest targets (cf.\ the left panel of Figure \ref{fig: observed planets heatmap}). Indeed, the reason fewer planets could be observed in these cases is that a handful of particularly hard-to-observe planets hogged a lot of the survey time. A detailed scheduling exercise would be needed to test whether the Gini coefficient of a target list correlates with scheduling difficulty.    

\section{Implications for Ariel}
We have shown that the Ariel survey leverage on certain axes of diversity can be increased by 10--25\% by breaking up the MCS into regular classes. It is harder to improve the leverage in multiple axes simultaneously, however: some axes will have better leverage than the one-room schoolhouse, and some worse. These gains and losses in leverage are on the order of 10\% with respect to the one-room schoolhouse, but the number of targets observed is $\sim$50\% lower (hurting other science cases), and the Gini coefficient is significantly higher (potentially a challenge for scheduling). It therefore seems likely that the one-room schoolhouse is close to optimal. In other words, if one is interested in survey leverage, then Ariel targets should be chosen based on their ease of observation: those with the greatest transit spectroscopy metric for the transit survey and those with the greatest emission spectroscopy metric for the emission survey \citep[][]{cowan_characterizing_2015,2017ApJ...844...27Z,2018PASP..130k4401K}.

Regardless of which class structures we considered, the number of planets observed increases by $\sim$50\% and the leverage increased by $\sim$30\% when including planet candidates in addition to confirmed planets (leverage computed in linear space doubled when including candidates).  This motivates the vetting and weighing of candidates before the Ariel target list is fixed in 2028. Planet confirmation efforts would benefit from guidance regarding which planet candidates to prioritize, as recently highlighted by \cite{2025arXiv250803801B}.

Compromises are necessary to maximize leverage in multiple axes of diversity.  Astronomers therefore urgently need to identify the axes of diversity most important for Ariel population-level studies.  
Once the axes of diversity have been defined, it is feasible to construct a target list that provides good leverage along each axis. In fact, the one-room schoolhouse is close to optimal and may be the overall best if science cases do not solely rely on survey leverage. This means that, to good approximation, the planet candidates that most urgently need to be vetted and weighed for Ariel are simply the easiest targets for atmospheric characterization.

\section*{Acknowledgments}
NBC acknowledges support from a Canada Research Chair, NSERC Discovery Grant, and McDonald Fellowship. He is also grateful to members of the Ariel Science Team and Ariel Consortium, who helped sharpen these ideas. The authors thank the Trottier Space Institute and l’Institut de recherche sur les exoplanètes for their financial support and dynamic intellectual environment. Giusi Micela, Emilie Panek, Robert Zellem, and two anonymous referees provided feedback that greatly improved the manuscript.

\bibliographystyle{aasjournal}

\bibliography{oja_template}

\begin{appendix}
\label{ap: lev}
In this appendix we show that the uncertainty on the slope of a trend line is inversely proportional to the leverage of the data. 
We write the coordinates for the $i^\text{th}$ datum:
\begin{equation}
    y_i = mx_i + b + \epsilon_i,
    \label{ap.eq: y}
\end{equation}
where $\epsilon_i$ is the error, and we assume independent errors for each datum. 
Using least-squares, the fitted slope $\hat m$ is
\begin{equation}
    \hat m = \frac{\sum_i (x_i - \bar x)(y_i - \bar y)}{\sum_i(x_i-\bar x)^2},
\end{equation} 
where $\bar y$ is the mean $y$-value.
Using Eqn \eqref{ap.eq: y} we find 
\begin{equation}
y_i - \bar y = m(x_i - \bar x) + \epsilon_i - \bar\epsilon.
\end{equation}
We can therefore rewrite the fitted slope $\hat m$ as
\begin{equation}
    \hat m = \frac{\sum_i \left( m(x_i - \bar x)^2 + (x_i - \bar x)(\epsilon_i - \bar\epsilon)\right)}{\sum_i(x_i-\bar x)^2} = m + \frac{\sum_i(\epsilon_i - \bar\epsilon)}{\sum_i(x_i-\bar x)}.
\end{equation}
The variance on the fitted slope is thus 
\begin{equation}
\begin{split}
    \sigma^2_{m} &= {\rm Var}(m) +{\rm Var}\left(\frac{\sum_{i}\epsilon_i}{\sum_i (x_i - \bar{x})}\right) + {\rm Var}\left(\frac{\sum_{i} \bar{\epsilon}}{\sum_{i} (x_i - \bar{x})}\right)= 0 + \frac{{\rm Var}(\sum_{i} \epsilon_i)}{\sum_{i} (x_i - \bar{x})^2} + 0 \\
    &= \frac{\sigma_y^2}{L^2},
\end{split}
\end{equation}
where $\sigma_y^2$ is the variance of the $y$-values with respect to the trend line, since we assume each $\epsilon_i$ to be independent. Taking the square root, we see that the uncertainty on the slope is inversely proportional to the leverage: $\sigma_m = \sigma_y/L$.

\end{appendix}

\end{document}